\documentstyle[amsmath,amssymb,12pt,epsf]{article}
\begin{document}
\begin{flushright} 
{CU-TP-1057}
\end{flushright}
\vskip50pt
\begin{center}
\begin{title}
\title{\large\bf The Energy Dependence of the Saturation Momentum}\footnote{This research is supported in part by the US Department of
Energy}\\

\vskip 10pt
{A.H. Mueller and D.N. Triantafyllopoulos\footnote{e-mail: dionysis@phys.columbia.edu}\\

Physics Department, Columbia University\\
New York, N.Y. 10027 USA}
\end{title}
\end{center}
\vskip 10pt
\begin{center}
{\bf Abstract}
\end{center}

We study BFKL evolution and, in particular, the energy dependence of the saturation momentum in the presence of saturation boundaries limiting
the region of linear BFKL evolution. In the case of fixed coupling evolution we confirm the previously found exponential term in $Q_s(Y)$ and
determine the prefactor $Y$ and $\alpha$ dependences.  In the running coupling case we find $Y^{1/6}$ corrections to the $Y^{1/2}$ exponential
behavior previously known.  Geometrical scaling of the scattering amplitude is valid in a wide range of momenta for fixed coupling evolution
and in a more restricted region for running coupling evolution.

\section{Introduction}

In 1983 Gribov, Levin and Ryskin\cite{Gri} introduced the idea of parton saturation in high energy hard scattering as a dual description of
unitarity. Since that time our understanding of saturation, and unitarity, in hard reactions has progressed considerably\cite{Ian}.  We now have
a simple model, the McLerran-Venugopalan model\cite{Mc,Jal,Kov}, which exhibits gluon saturation in a simple and, likely, fairly general
manner. This model is now being used in order to understand general features of heavy ion reactions\cite{Kha,Lev,Nar,Kra,Bai}. In deep
inelastic scattering the Golec-Biernat and W\"usthoff model incorporates the essential elements of saturation and gives a surprisingly good fit
to much of the HERA data for $F_2$ and for diffractive production at small values of $x$\cite{Gol,Sta}.  There remain, however, many
uncertainties in our understanding and application of saturation ideas.

The points we wish to address in this paper are the value and energy dependence of the saturation momentum and the form of the scattering
amplitude on the perturbative side of the saturation line.  To be more specific, and in order to illustrate the issues, consider the
scattering of a QCD dipole of size $1/Q$ on either a hadron or on another dipole of size $1/\mu$ and with relative rapidity $Y.$  Then the
saturation momentum, $Q_s(Y)$, is the momentum at which determining the scattering changes from a purely perturbative problem, for $Q>Q_s,$ to
a nonperturbative but weak coupling problem where unitarity limits have been reached, for $Q<Q_s.$

The main question which naturally arises is the $Y$-dependence of $Q_s$ and what information (dynamics) is actually necessary to
control in order to calculate the $Y-$dependence of $Q_s.$  In general terms one expects BFKL\cite{Lip,Sov} dynamics, but not necessarily the
BFKL saddle-point solution, to be the relevant dynamics since this is the evolution which leads to high density partonic systems.  Of course
one cannot expect linear BFKL evolution to be accurate when $Q\lesssim Q_s(Y).$  As a rough guess one can use the BFKL saddle-point solution for
high energy scattering at large $Q$ and then define $Q_s(Y)$ as the value at which this scattering amplitude reaches its unitarity bound.  This
was done in Ref.15 for fixed coupling BFKL evolution with the result $\ln(Q_s^2(Y)/\Lambda^2) = {2\alpha N_c\over
\pi}\ {\chi(\lambda_0)\over 1-\lambda_0}Y$ where $\lambda_0$ is the solution to $\chi^\prime(\lambda_0)(1-\lambda_0) = - \chi(\lambda_0)$ with
$\chi$ the usual BFKL eigenvalue function.  The danger with this procedure is that one cannot justify using the saddle-point method for solving
asymptotic BFKL evolution even when the scattering amplitude is small.  (We shall show in Secs.6 and 7 that this procedure gives the
exponential parts of the $Y$-dependence of $Q_s$ correctly, but misses $Y$-dependent prefactors.)  This discussion was extended in Ref.16 to the
running coupling case where it was also observed that one can expect the scattering amplitude to be a function of $Q^2/Q_s^2$ in an extended
region where
$Q^2/Q_s^2>1.$  In Ref.17 a numerical study of the Kovchegov equation\cite{gov} was carried out.  This has the advantage that BFKL dynamics is
used when $Q/Q_s{\gg}1$ while unitarity is imposed in a realistic way when $Q/Q_s\lesssim 1.$  The energy dependence found numerically for $Q_s$ is
close to that expected from simple saddle-point BFKL dynamics in the region $Q/Q_s>1.$

In this paper we give a new procedure to solving linear BFKL dynamics in the region $Q/Q_s> 1$ in such a way that the matching with the
nonlinear dynamics present when $Q/Q_s \lesssim 1$ should be smooth.  To illustrate the main problem we face here consider the scattering of a
dipole of size $1/Q$ on a dipole of size $1/\mu$ in fixed (weak) coupling BFKL evolution.  Suppose we are given $Q_s(Y)$ and that $Q/Q_s\gg 1.$ 
Can we expect the scattering amplitude to be given by the saddle-point approximation to BFKL evolution?  The answer is no, not in general. 
Because of diffusion there will be many paths, in the functional integral sense, which go from $\mu$ at $Y=0$ to $Q$ at $Y$ and on the way pass
through the saturation region.  These paths should not be allowed in the true solution to the scattering problem.   This, however, seems to be
the only difficulty with using the saddle-point approximation.  We face this difficulty by converting the usual diffusive behavior in BFKL
dynamics to one with an absorbing boundary near $Q_s(Y),$ that is we throw away all paths which go into the saturation region.

In order to gain confidence that diffusion with an absorptive barrier is the right thing to do, we first study a problem whose answer is known
and which has many similarities to evolution in the presence of saturation, namely non-forward scattering in BFKL evolution.  It is well-known
that scattering at a non-zero momentum transfer 
$q$ cuts off all infrared dynamics below $q.$  In scattering a dipole of size  $x$ on a dipole of size $x_0,$ with $qx, qx_0\ll 1$ we expect
that one can get the correct, non-forward, BFKL behavior by using forward BFKL evolution but with an absorptive boundary\cite{Sov} which
eliminates diffusive paths that go into the momenta region below  $q.$  In Secs. 4 and 5 we verify that this is the case.

In Sec.6 we evaluate BFKL evolution in the presence of saturation in the case of dipole-dipole scattering and where the coupling is fixed.  Our
main results are given in (47) for the scattering amplitude, and in (48) for the saturation momentum.  The exponent in (48) is as previously
found\cite{ler,ura} while the $\alpha$ and $Y-$dependence of the prefactors is new.  Eq.(47) exhibits geometric scaling\cite{Sta} when $\ln
(Q^2/Q_s^2) \gtrsim 1,$ and what is perhaps even more remarkable is that there are no unknown prefactors in $T.$  Eq.47 is valid so long as
$\ln^2(Q^2/Q_s^2)\ll {4\alpha N_c\over \pi} \chi^{\prime\prime}(\lambda_0)Y,$ that is within the diffusion regime for BFKL evolution in the
absence of boundaries.

In Sec.7, we deal with the running coupling case.  This discussion should apply to high-energy hard scattering on protons.  Our main results
are contained in (83), (84) and (85).  In (83) we find, for $\rho_s=\ln (Q_s^2/\Lambda^2),$ the expected leading ${\sqrt{Y}}$
behavior\cite{ura} along with a $Y^{1/6}$ correction while (84) and (85) exhibit geometric scaling, but now only for $\ln (Q^2/Q_s^2) \lesssim
[{N_c\chi^{\prime\prime}(\lambda_0)\over \pi b(1-\lambda_0)\chi(\lambda_0)}Y]^{1/6}.$  What is remarkable is that $Q_s$ has no knowledge
of the target, whether that target be a hadron or a small dipole, so long as $Y$ is in the asymptotic regime.  For a very small dipole target,
of size $1/\mu,$ the asymptotic regime only begins when $Y \gtrsim {\pi(1-\lambda_0)\over 2bN_c\chi(\lambda_0)\alpha^2(\mu)}.$  In the running
coupling case we are unable to determine target-dependent prefactors in  $T$  either when the target is a hadron and even when it is a small
dipole.

\section{BFKL Evolution; Naive View}

We consider the forward scattering amplitude for a dipole of size $1/Q$ on a dipole of size $1/\mu.$  In the large $N_c$ limit and at the
leading logarithmic level, this is described by BFKL evolution.  When the strong coupling $\alpha$ is fixed, the amplitude is

\begin{equation}
T(Q,\mu, Y) = {\pi\alpha^2\over \mu^2} \int {d\lambda\over 2\pi i}\ {1\over \lambda^2(1-\lambda)^2}
\exp\left[{2\alpha N_c\over \pi}\
\chi(\lambda)Y-(1-\lambda) \ln {Q^2\over \mu^2}\right],
\end{equation}

\noindent where

\begin{equation}
\chi(\lambda) = \psi(1) - {1\over 2} \psi(\lambda) - {1\over 2} \psi (1-\lambda),
\end{equation}

\noindent with $\psi(\lambda) = \Gamma^\prime(\lambda)/\Gamma(\lambda)$ as usual and the integation contour being parallel to the imaginary
axis with $O<Re(\lambda) < 1.\   Y$ is the relative rapidity of the two objects.  Expression (1) is symmetric under the interchange $Q
\leftrightarrow \mu,$ and is normalized at $Y=0$ to be the elementary cross section of two dipoles at the two gluon exchange approximation,
which is

\begin{equation}
T(Q, \mu, Y=0) = {2\pi\alpha^2\over Q_>^2} \left(1 + \ln {Q_>\over Q_<}\right).
\end{equation} 

\noindent In the above, $Q_>=\max (Q,\mu)$ and $Q_<=\min (Q,\mu).$

Now we define a particular line in the $\ln (Q^2/\mu^2) - Y$ plane by the following two conditions

\begin{equation}
{2\alpha N_c\over \pi} \chi^\prime(\lambda_0)Y + \ln {Q_0^2\over \mu^2} = 0,
\end{equation}

\noindent and

\begin{equation}
{2\alpha N_c\over \pi} \chi(\lambda_0) Y - (1-\lambda_0) \ln {Q_0^2\over \mu^2} = 0.
\end{equation}

\noindent Eq.(4) is just the saddle-point condition, while Eq.(5) means that along the line $Q^2=Q_0^2(Y)$ the exponent in (1) at the
saddle-point vanishes.  As we shall see in a while, this is a line of almost, but not exactly, constant amplitude.  The solution to (4) and (5)
is

\begin{equation}
{\chi^\prime(\lambda_0)\over \chi(\lambda_0)} = - {1\over 1-\lambda_0},
\end{equation}

\begin{equation}
Q_0^2(Y) = \mu^2\exp\left({2\alpha N_c\over \pi}\ {\chi(\lambda_0)\over 1-\lambda_0}Y\right).
\end{equation}

\noindent The graphical solution to (6) is shown in Fig.1.

\begin{center}
\begin{figure}[htb]
\epsfbox[0 0 169 151]{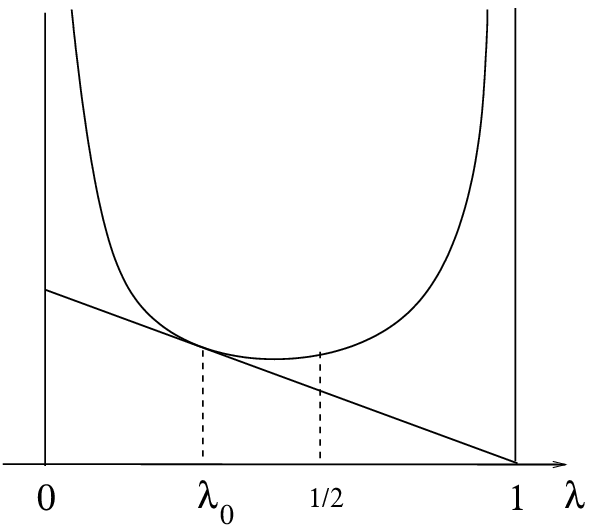}
\centerline{Fig.1}
\end{figure}
\end{center}

The value of $\lambda_0,$ as determined by (6), is $\lambda_0=0.372,$ which is not too far from $1/2,$ so that $Y-$evolution is the dominant
one in the process and the approach is reasonable.  One can now evaluate the scattering amplitude in a region around $Q_0^2(Y),$ by the saddle
point method, when $\alpha Y$ is large.  Since $\lambda_0$ is chosen to satisfy the saddle-point condition, we expand $\chi(\lambda)$ around
this point.  Then by using (4) and (5), the amplitude becomes

\begin{align}
T=& {\pi\alpha^2\over \lambda_0^2(1-\lambda_0)^2\mu^2} \left[{Q_0^2(Y)\over Q^2}\right]^{1-\lambda_0}\nonumber\\
& \times  \int {d\lambda\over 2\pi i}\exp\left[{\alpha N_c\over \pi}\chi^{\prime\prime}(\lambda_0) Y (\lambda-\lambda_0)^2 + (\lambda-\lambda_0) \ln
{Q^2\over Q_0^2(Y)}\right].
\end{align}

\noindent The Gaussian integral is easily done and we obtain

\begin{equation}
T={\pi\alpha^2\over \lambda_0^2(1-\lambda_0)^2\mu^2}\ {1\over{\sqrt{4\alpha N_c\chi^{\prime\prime}(\lambda_0)Y}}}\left[{Q_0^2(Y)\over
Q^2}\right]^{1-\lambda_0}\exp\left[-{\pi \ln^2(Q^2/Q_0^2(Y))\over 4\alpha N_c\chi^{\prime\prime}(\lambda_0)Y}\right].
\end{equation}

\noindent Since higher derivatives of $\chi(\lambda)$ have been neglected, this result is valid so long as $\vert \ln (Q^2/Q_0^2)\vert \ll
4\alpha N_c\chi^{\prime\prime}(\lambda_0)Y/\pi.$

A couple of comments need to follow here.  (i)  When the dipole size is inside the diffusion region $\ln^2(Q^2/Q_0^2) \ll
4\alpha\chi^{\prime\prime}(\lambda_0) Y/\pi,$ and ignoring, for the moment, the slowly varying prefactor $1/{\sqrt{\alpha Y}},$ the dominant
factor of the amplitude is $(Q_0^2/Q^2)^{1-\lambda_0}.$  This of course has a scaling form, with momentum scale $Q_0^2(Y)$ as given by (7); the
$Y-$dependence of this scale is exponential and the coefficient in the exponent is known.  (ii) The amplitude along the line $Q^2=Q_0^2(Y),$ as
claimed earlier, is close to being constant, but not quite as it behaves as $1/{\sqrt{\alpha Y}}.$  We are going to find lines of constant
amplitude later, but an important thing to notice here is, that one can evolve the system to arbitrary large rapidity along this line,
without facing the problem of making the amplitude too big and violating unitarity constraints.  This is happening because as we increase $Y,$
at the same time we exponentially suppress the dipole size.  This is in sharp contrast to the usual BFKL evolution, where one considers final
dipole sizes comparable to the initial one, or more precisely inside its diffusion radius $\ln^2(Q^2/\mu^2) \ll 4\alpha
N_c\chi^{\prime\prime}(1/2)Y/\pi.$

\section{Diffusion into the Saturation Regime}

The analysis so far is naive, in the sense that we have ignored problems arising from the diffusion of the solution, as given in (9), into the
saturation regime.  When the forward scattering amplitude becomes of order $2\pi/\mu^2$ (or equivalently the amplitude in impact parameter space
of order 1), one should not trust the solution any more, since unitarity constraints are violated.  Therefore the dynamics in that region
cannot be represented by the BFKL evolution.  We do not intend to find the amplitude in this saturation regime, where presumably we would
have to solve the (non-linear) Balitsky-Kovchegov equation\cite{gov,lit}, but the problem is that the solution in the purely perturbative
region is altered because of unitarity effects.  At this level, one can see from (9) that the amplitude becomes of order $2\pi/\mu^2,$ when the
dipole size $1/Q$ becomes $\ln (Q^2/Q_0^2) \approx - \ln ({\sqrt{\alpha Y}}/\alpha^2)/(1-\lambda_0).$  This logarithmic distance
of
$Q^2$ from $Q_0^2(Y)$ is small compared to the diffusion radius ${\sqrt{\alpha Y}}$ when $\alpha Y$ is large.  Thus, even though we start with
an initial condition that the dipole size is in the perturbative region, as $Y$ increases the diffusion drives the dipoles to bigger sizes that
enter the saturation regime and therefore it will, partly, invalidate the result as given in (9).

This problem caused by diffusion can be seen from a mathematical point of view, when one tries to reproduce (9) by doing two successive
evolutions in rapidity. Imagine first that we evolve the amplitude from zero rapidity to $Y/2.$  Then assume that the amplitude at this
rapidity is given by (9), with $Y \to Y/2,$ for dipole sizes $1/Q$ such that the amplitude is less or equal to $2\pi/\mu^2,$ and given by
$2\pi/\mu^2$ for larger dipoles.  This serves as an initial distribution at $Y/2$ in a simple way to \underline{impose} unitarity.  Then one can
evolve to find the amplitude at rapidity $Y$ always in the perturbative regime.  This is a straightforward calculation that we don't present
here and one can find that the result agrees with (9) provided that ${\sqrt{\alpha Y}} \lesssim \ln ({\sqrt{\alpha Y}}/\alpha^2).$  This is of
course a small, for our purposes, evolution in $Y.$  It simply states the fact, that the solution is incorrect when the rapidity is large
enough so that the dipoles start diffusing into the saturation regime.

We shall come back to resolve this issue by imposing unitarity in a more proper way in section 6.  Before this, and in order to motivate the
work in that section, we find it useful to consider the non-forward scattering in section 4 and how this is related to the diffusion equation
in the presence of an absorptive barrier in section 5.

\section{Non-Zero Momentum Transfer}

Just for the purposes of this section and  the following one, we adopt a slightly different notation.  We consider now the scattering of a
dipole of size $x$ on a dipole of size $x_0,$  when the momentum transfer of the scattering is q.  In this case the amplitude is

\begin{equation}
T_{nf}(x, x_0, Y,q) = {\pi\alpha^2xx_0\over 2} \int {d\nu\over 2\pi}\ {1\over (\nu^2+{1\over 4})^2}
E_q^{0\nu\ast}(x_0)E_q^{0\nu}(x)\exp[{2\alpha N_c\over \pi}\chi(\nu)Y],
\end{equation}

\noindent where

\begin{equation}
\chi(\nu) = \psi(1) - {1\over 2} \psi({1\over 2} + i\nu) - {1\over 2} \psi({1\over 2} - i\nu),
\end{equation}

\noindent and

\begin{equation}
E_q^{0\nu}(x) = {2i\nu\over \pi} x^{2i\nu}\int d^2R {e^{i\vec{q}\cdot \vec{R}}\over (\vert \vec{R} + {\vec{x}\over 2}\vert \vert\vec{R} -
{\vec{x}\over 2}\vert)^{1+2i\nu}}.
\end{equation}

\noindent Following the method presented in [21,22], we calculate the amplitude in the Appendix when $qx,  qx_0\ll 1.$  We obtain (for large
$Y$)
\begin{align}
T_{nf}= & 8\pi\alpha^2xx_0{1\over {\sqrt{\pi DY}}}e^{(\alpha_P-1)Y}\nonumber\\
& \times \left\{\exp\bigg[-{\ln^2(x/x_0)^2\over DY}\right]-\exp\left[-{\ln^2(cq^2xx_0)^2\over DY}\bigg]\right\},
\end{align}

\noindent where $\alpha_P-1=4\alpha N_c(\ln 2)/\pi, D = 56\alpha N_c\zeta(3)/\pi,$ the constant appearing is $c=e^{2\gamma}/64$ and $\gamma
= 0.577... .$  The result is valid in the domain $\vert \ln (x/x_0)^2\vert \ll DY.$  Here we have done the standard BFKL evolution by
expanding the $\chi$  function around its extremum at $\nu=0$ (corresponding to $\lambda = 1/2$ in the previous notation).  Eq.(13) can also be
derived by looking at the scattering at a fixed impact parameter  $b$  and then taking its Fourier tansformation with the integration limits
for $b$  being from $\sim \max (x,x_0)$ to $\sim 1/q.$

The first term in the curly bracket of (13) can be recognized as the one appearing in the forward case.  Notice that the second term has the
same diffusion pattern and, because of the minus sign, it will eventually cut off large dipole sizes. We will return to comment on this in the
next section.

\section{Diffusion in the Presence of an Absorptive Barrier}

Going back to (13) and looking at the two terms in the bracket accompanied by the $1/{\sqrt{\pi DY}}$ prefactor, we see that they obey a
diffusion equation.  What is not so clear, for the moment, is the physical interpretation of the second term in the diffusion mechanisn, a task
that we now turn into.  For reasons that will be apparent soon we review the diffusion equation in the presence of an absorptive barrier, which
is

\begin{equation}
{\partial\psi(\rho,t)\over \partial t} = {1\over 4} {\partial^2\psi(\rho,t)\over \partial t^2},
\end{equation}

\noindent with the boundary condition

\begin{equation}
\psi(-\rho_a,t) = 0.
\end{equation}

\noindent We convert this into a Green's function problem,

\begin{equation}
\psi(\rho,t) = \int_{-\infty}^\infty d\rho^\prime G(\rho,\rho^\prime,t-t^\prime) \psi(\rho^\prime,t^\prime),
\end{equation}

\noindent where the kernel should satisfy (for $\rho,\rho^\prime\geq - \rho_a$)

\begin{equation}
G(\rho,\rho^\prime,0) = \delta(\rho-\rho^\prime).
\end{equation}

\noindent If we define the Laplace transformation of $G(\rho,\rho^\prime,t-t^\prime)$ with respect to time by

\begin{equation}
G(\rho,\rho^\prime,t-t^\prime) = \int {d\omega\over 2\pi i} e^{\omega(t-t^\prime)}G_\omega(\rho,\rho^\prime),
\end{equation}

\noindent then Eqs.(14) and (17) imply

\begin{equation}
{\partial^2G_\omega(\rho,\rho^\prime)\over \partial\rho^2} - 4\omega G_\omega(\rho,\rho^\prime)= - 4\delta(\rho-\rho^\prime).
\end{equation}

\noindent The solution to this equation, satisfying the condition $G_\omega(-\rho_a,\rho^\prime) = 0,$ is (for $\rho,\rho^\prime \geq -
\rho_a$)

\begin{equation}
G_\omega(\rho,\rho^\prime) = {1\over{\sqrt{\omega}}}\left[e^{-2{\sqrt{\omega}}\vert\rho-\rho^\prime\vert}-e^{-2{\sqrt{\omega}}(2\rho_a+ \rho +
\rho^\prime)}\right].
\end{equation}

\noindent Finally, we perform the integration in  (18) along the imaginary axis, to arrive at

\begin{equation}
G(\rho,\rho^\prime,t-t^\prime) = {1\over{\sqrt{\pi t}}}\left[e^{-{(\rho-\rho^\prime)^2\over t-t^\prime}}-e^{-{(2\rho_a + \rho +
\rho^\prime)^2\over t-t^\prime}}\right].
\end{equation}

\noindent Eq.16 is quite general and by choosing an initial condition $\psi(\rho^\prime,t^\prime=0) = \delta(\rho^\prime),$ which is the 
relevant one for our purposes, we obtain

\begin{equation}
\psi(\rho,t) = {1\over {\sqrt{\pi t}}} \left(e^{-{\rho^2\over t}}-e^{-{(2\rho_a+\rho)^2\over t}}\right).
\end{equation}

\noindent If we let $t \to DY, \rho \to \ln (x_0^2/x^2)$ and $\rho_a\to - \ln (c x_0^2q^2),$ then $\psi$ becomes

\begin{equation}
\psi(x,Y) = {1\over {\sqrt{\pi DY}}} 
\left\{\exp\left[-{\ln^2(x/x_0)^2\over DY}\right] -\exp\left[-{\ln^2(cq^2x x_0)^2\over DY}\right]\right\},
\end{equation}

\noindent which is, apart from the exponential increase and the $8\pi\alpha^2x x_0$ prefactor, identical to (13).

Here we have two equivalent descriptions.  Say that we want to calculate the forward scattering amplitude with momenta (inverse dipole size) $Q
\lesssim q$ cut-off, where  $q$  is a fixed momentum.  An intuitive way is to look at the non-forward amplitude at momentum transfer  $q.$ 
This is the only new scale entering the problem and thus will eventually offer the infrared cut-off scale in the problem.  An alternative
approach is to look directly at forward scattering, but at the same time do not allow diffusion, in the dipole size, to go into the region $Q
\lesssim q.$  The last must happen in an absorptive way; if the dipole hits the boundary at $q,$  it never comes back in the region $Q\gtrsim q$ as we
show in Fig.2.

\begin{center}
\begin{figure}
\epsfbox[0 0 344 212]{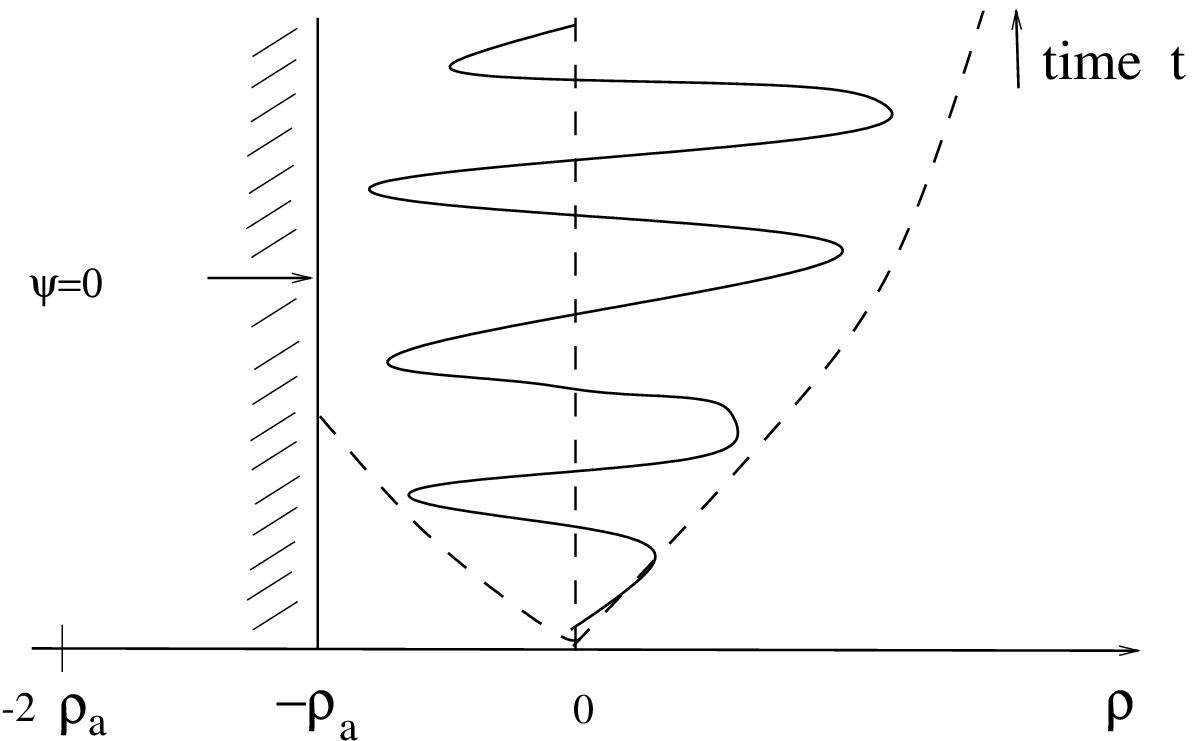}
\centerline{Fig.2}
\end{figure}
\end{center}

Going back to Eqs.(22), (23) one can expand the exponentials in the diffusion region $\rho^2,\rho_a^2 \ll t.$  Then (22) becomes

\begin{equation}
\psi(\rho,t) \simeq {4\rho_a\over {\sqrt{\pi}}}\ {(\rho + \rho_a)\over t^{3/2}}.
\end{equation}

\noindent Notice the $t^{3/2}$ power in the denominator. It has the interpretation that the probability of not diffusing into $Q \lesssim q,$  by
cutting out these paths, is $1/\alpha Y$ times the probability of all paths.

We apply these ideas in the next section, where we return to impose unitarity effects in the BFKL evolution considered in Section 2.

\section{BFKL Evolution in the Presence of Saturation}

Referring to the starting equation (1) for the forward amplitude, we define a new line $Q_c^2(Y),$ which is close to the line $Q_0^2(Y)$ in
the $\ln(Q^2/\mu^2) - Y$ plane, by the following conditions

\begin{equation}
{2\alpha N_c\over \pi} \chi^\prime(\lambda_c) Y + \ln {Q_c^2\over \mu^2} = 0
\end{equation} 

\noindent and

\begin{equation}
{2\alpha N_c\over \pi} \chi(\lambda_c)Y - (1-\lambda_c) \ln {Q_c^2\over \mu^2} = {3\over 2} \ln[{4\alpha
N_c\chi^{\prime\prime}(\lambda_c)Y\over \pi}].
\end{equation}

\noindent Eq.(25) is again a saddle-point condition, as Eq.(4), while (26) has an extra $(3/2) \ln (\alpha Y)$ term compared to (5).  This
definition of the critical line $Q_c^2(Y)$, will result in an extra $(\alpha Y)^{3/2}$ factor in the amplitude, which will cancel the $(\alpha
Y)^{3/2}$ factor that is anticipated from the discussion of the previous section, more precisely from (24).  Therefore we will be able to
recover lines of constant amplitude.  The solution to (25) and (26) is

\begin{equation}
(1-\lambda_c)\chi^\prime(\lambda_c) + \chi(\lambda_c) = 
{3\ln[{4\alpha N_c\chi^{\prime\prime}(\lambda_c)Y\over \pi}]\over {4\alpha N_c\over\pi}Y},
\end{equation}

\noindent and

\begin{equation}
Q_c^2(Y) = {\mu^2 \exp[{2\alpha N_c\over \pi}\ {\chi(\lambda_c)\over 1-\lambda_c}Y]\over [{4\alpha N_c\over \pi}
\chi^{\prime\prime}(\lambda_c)Y]^{3\over 2(1-\lambda_c)}}.
\end{equation}

\noindent Eq.(27) determines $\lambda_c$ which is a function of the rapidity  $Y,$ and not a pure number like $\lambda_0.$  However, when
$\alpha Y$ is large, $\lambda_c$ is very close to  $\lambda_0.$  In this case (27) reduces to

\begin{equation}
\lambda_c-\lambda_0 = {3\ln[{4\alpha N_c\over \pi}\chi^{\prime\prime}(\lambda_0)Y]\over (1-\lambda_0){4\alpha N_c\over
\pi}\chi^{\prime\prime}(\lambda_0)Y}.
\end{equation}

\noindent In the exponential in (28) we notice that

\begin{equation}
{\chi(\lambda_c)\over 1-\lambda_c} = {\chi(\lambda_0)\over 1-\lambda_0} + {1\over 2}\  {\chi^{\prime\prime}(\lambda_0)\over
1-\lambda_0}(\lambda_c-\lambda_0)^2 + \cdot \cdot \cdot,
\end{equation}

\noindent as the linear term cancels when we make use of (6).  This exponential can now be evaluated at $\lambda_0,$ since (29) and (30) imply
that the remaining term is of order $\ln^2(\alpha Y)/\alpha Y$ which is small.  Then Eq.(28) becomes

\begin{equation}
Q_c^2(Y) = {\mu^2 \exp[{2\alpha N_c\over \pi}\ {\chi(\lambda_0)\over 1-\lambda_0}Y]\over [{4\alpha N_c\over \pi}
\chi^{\prime\prime}(\lambda_0)Y]^{{3\over 2(1-\lambda_0)}}}={Q_0^2(Y)\over [{4\alpha N_c\over \pi}\chi^{\prime\prime}(\lambda_0)Y]^{{3\over
2(1-\lambda_0)}}}.
\end{equation}

\noindent It is clear that the lines $Q_0^2(Y)$ and $Q_c^2(Y)$ are close in the $\ln (Q^2/\mu^2) - Y$ plane.

Let's call $E$ the exponent in (1).  Expanding $\chi(\lambda)$ around $\lambda_c,$ this exponent takes the form

\begin{displaymath}
E = {2\alpha N_c\over \pi} Y [\chi(\lambda_c) + (\lambda - \lambda_c) \chi^\prime(\lambda_c) + {1\over 2} (\lambda -
\lambda_c)^2\chi^{\prime\prime}(\lambda_c) + \cdot\cdot\cdot]
\end{displaymath}
\begin{equation}
-[(1-\lambda_c) -(\lambda - \lambda_c)]\ln {Q_c^2\over \mu^2} - [(1-\lambda_c) - (\lambda - \lambda_c)]\ln {Q^2\over Q_c^2}.
\end{equation}

\noindent Making use of the definitions of $\lambda_c$ and $Q_c^2(Y),$ which are (25) and (26), the last expression simplifies to

\begin{displaymath}
E = {1\over 2}\  {2\alpha N_c\over \pi} \chi^{\prime\prime}(\lambda_c)Y(\lambda-\lambda_c)^2 + (\lambda - \lambda_c) \ln
{Q^2\over Q_c^2}
\end{displaymath}
\begin{equation}
+ {3\over 2} \ln\left[{4\alpha N_c\chi^{\prime\prime}(\lambda_c)Y\over \pi}\right]-(1-\lambda_c) \ln {Q^2\over Q_c^2},
\end{equation}

\noindent where the last two terms are independent of the integration variable $\lambda.$  Once again we do the Gaussian integration to obtain

\begin{displaymath}
T={\pi\alpha^2\over \lambda_c^2(1-\lambda_c)^2\mu^2}\left[{4\alpha N_c\chi^{\prime\prime}(\lambda_c)Y\over \pi}\right]^{3/2}\left({Q_c^2\over
Q^2}\right)^{1-\lambda_c}
\end{displaymath}
\begin{equation}
\times {1\over {\sqrt{4\alpha N_c\chi^{\prime\prime}(\lambda_c)Y}}} \exp\left[-{\pi \ln^2(Q^2/Q_c^2)\over 4\alpha
N_c\chi^{\prime\prime}(\lambda_c)Y}\right].
\end{equation}

\noindent To simplify the notation we define

\begin{equation}
\rho = \ln {Q^2\over \mu^2}, \rho_c = \ln {Q_c^2\over \mu^2},
\end{equation}

\noindent and

\begin{equation}
t={4\alpha N_c\chi^{\prime\prime}(\lambda_c)Y\over \pi}.
\end{equation}

\noindent  Then the amplitude in terms of $\rho$ and $t$ becomes

\begin{equation}
T(\rho,t) = {\pi\alpha^2\over \lambda_c^2(1-\lambda_c)^2\mu^2} e^{-(1-\lambda_c)(\rho-\rho_c)}t^{3/2}\psi(\rho-\rho_c,t),
\end{equation}

\noindent where we defined

\begin{equation}
\psi(\rho-\rho_c,t) = {1\over{\sqrt{\pi t}}} e^{-(\rho-\rho_c)^2\over t}.
\end{equation}

\noindent It is obvious that $\psi$ represents the diffusive part of the amplitude since it satisfies

\begin{equation}
{\partial\over \partial t} \psi (\rho-\rho_c,t) = {1\over 4} {\partial^2\over \partial\rho^2} \psi (\rho-\rho_c,t).
\end{equation}

It is at this point that unitarity of the amplitude has to be imposed.  Eq.(38) is not the proper solution of (39) in the presence of
saturation.  When solving this diffusion equation, we require that we do not include paths which go into the saturation region; momenta $Q
\lesssim Q_s,$  where $Q_s$ is the saturation boundary, will be cut out.  We do this, following the discussion in the two previous sections, by
requiring that $\psi$ vanish very close to the saturation boundary.  Of course in the problem we consider before, the boundary was
$Y-$independent, since the fixed momentum transfer $q$ was the infrared cutoff.  Here $Q_s$ is $Y-$dependent and it will be a line parallel to
the critical one $Q_c.$  However, Eq.(39) is in terms of $\rho-\rho_c$ and $t$ variables and we can put a time independent absorptive barrier
in $\rho-\rho_c,$ which corresponds to a $Y-$dependent barrier in momentum  $Q,$ thus making our approach for cutting momenta $Q \lesssim Q_s(Y)$
reasonable.

We still have to exhibit the above in detail and we start by considering

\begin{equation}
\psi_s(\rho,t) = \psi(\rho-\rho_c,t) - \psi(\rho-\rho_c+ 2\Delta,t),
\end{equation}

\noindent where $\psi$ is the one given by (38) (the solution with no boundary conditions), while in $\psi_s(\rho,t)$ we require $\rho\geq
\rho_c -\Delta,$ as we note that $\psi_s=0$ when $\rho = \rho_c-\Delta.$  $\Delta$ is a parameter that has to be determined.  Eq.37 will now
become
\begin{equation}
T={\pi\alpha^2\over \lambda_c^2(1-\lambda_c)^2\mu^2}e^{-(1-\lambda_c)(\rho-\rho_c)}t^{3/2}{1\over{\sqrt{\pi t}}}[e^{-{(\rho-\rho_c)^2\over
t}}-e^{-{(\rho-\rho_c+2\Delta)^2}\over t}].
\end{equation}

\noindent When $\rho-\rho_c$ and $\Delta$ are much smaller than the diffusion radius ${\sqrt{t}},$ the t-dependent prefactors will cancel with
the $1/t$ factor coming from the expansion of the exponentials.  In this region we can also replace $\lambda_c$ by $\lambda_0,$  since
$(\lambda_c-\lambda_0)(\rho-\rho_c)\ll (\ln  t)/{\sqrt{t}}\ll 1$ as implied by (29).  Then we are lead to

\begin{equation}
T={\pi\alpha^2\over \lambda_0^2(1-\lambda_0)^2\mu^2} {4\over {\sqrt{\pi}}}\Delta(\rho-\rho_c+\Delta) e^{-(1-\lambda_0)(\rho-\rho_c)}.
\end{equation}

\noindent This expression becomes maximum at the point

\begin{equation}
\rho_s=\rho_c-\Delta + {1\over 1-\lambda_0},
\end{equation}

\noindent which is a finite and $\alpha$-independent distance away from the point where it becomes zero, as shown in Fig.3.  Considering the
amplitude at $\rho_s,$ we determine $\Delta$ by setting

\begin{center}
\begin{figure}
\epsfbox[0 0 390 195]{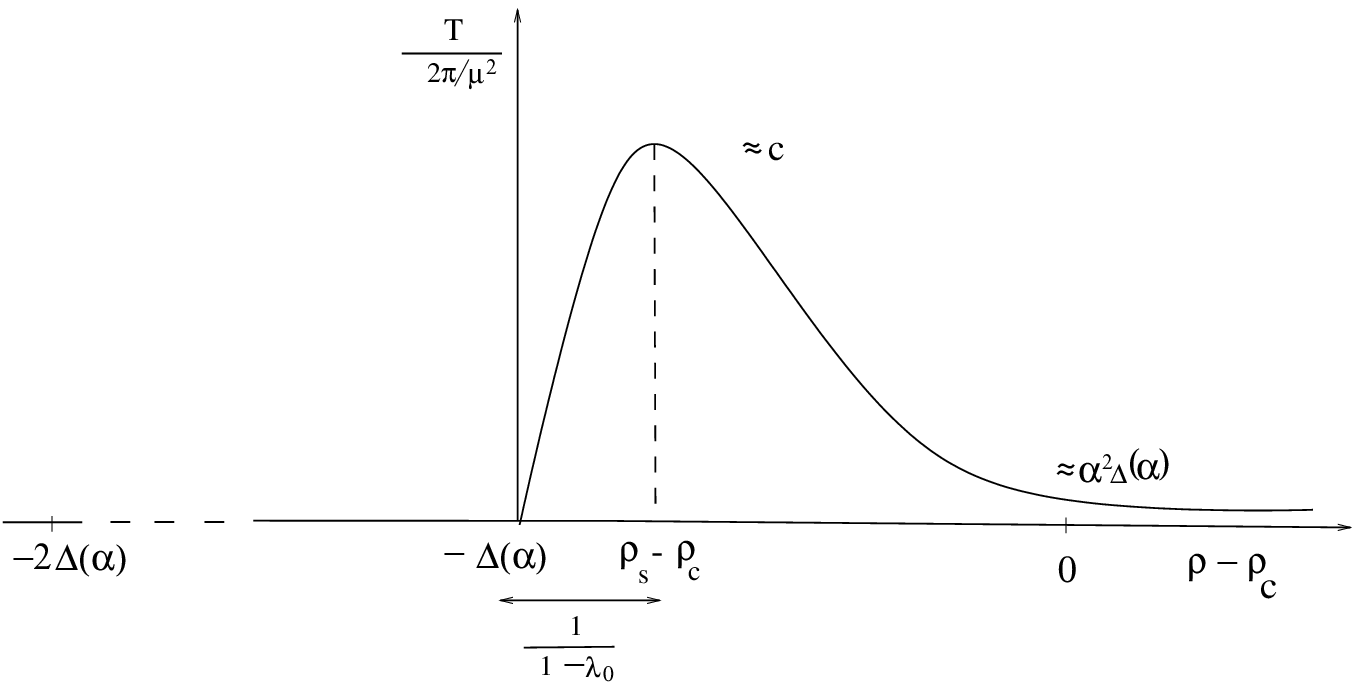}
\centerline{Fig.3}
\end{figure}
\end{center}

\begin{equation}
T(\rho_s,t) = {2\pi c\over \mu^2},
\end{equation}

\noindent with  $c$  a constant of order $1$  as required by unitarity.  Eqs.(42), (43) and (44) give

\begin{equation}
\alpha^2\Delta e^{(1-\lambda_0)\Delta}={ce{\sqrt{\pi}}\lambda_0^2(1-\lambda_0)^3\over 2}.
\end{equation}

\noindent We can solve this transcendental equation by iteration, for small coupling $\alpha,$ and the solution is

\begin{equation}
\Delta(\alpha) = {2\over 1-\lambda_0}\ln {1\over \alpha} - {1\over 1-\lambda_0} \ln \ln {1\over \alpha} + O({\rm const}),
\end{equation}

\noindent where all the constants appearing in the right-hand side of (45) have been absorbed in the (irrelevant) constant term in (46).

We are finally in a position to give a result for the forward scattering amplitude, and switching back to our original notation we have

\begin{equation}
T= {2\pi\over \mu^2}{2 \alpha^2\Delta(\alpha)\over {\sqrt{\pi}}\lambda_0^2(1-\lambda_0)^2}\left[\ln {Q^2\over Q_c^2(Y)} +
\Delta(\alpha)\right]\left[{Q_c^2(Y)\over Q^2}\right]^{^{1-\lambda_0}},
\end{equation}

\noindent with $Q_c^2(Y)$  given by (31) and $\Delta(\alpha)$ by (46).

This result is valid in the diffusion region $\ln^2(Q^2/Q_c^2) \ll 4\alpha N_c\chi^{\prime\prime}(\lambda_0)Y/\pi$ when
$Q\geq Q_c$ (to the right of the critical line in Fig.4) and in the region $\vert \ln(Q^2/Q_c^2)\vert \ll \Delta(\alpha)$ when $Q\leq Q_c$
(to the left of the critical line).  The validity of (47) also requires the rapidity to be large enough, so that

\begin{center}
\begin{figure}
\epsfbox[0 0 247 215]{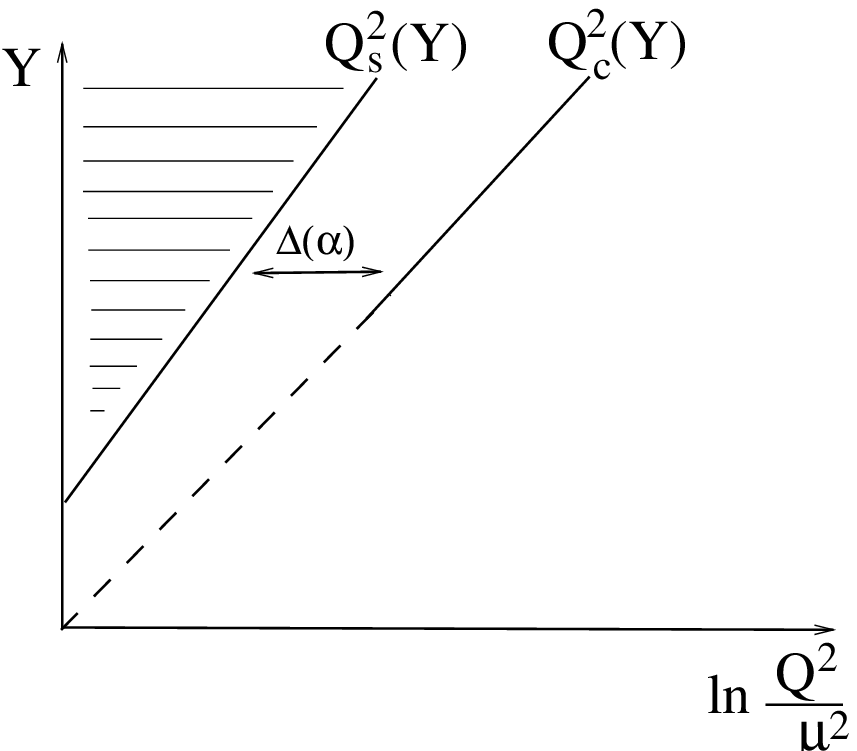}
\centerline{Fig.4}
\end{figure}
\end{center}

\begin{displaymath}
Y \gg {4\pi\over (1-\lambda_0)^2N_c\chi^{\prime\prime}(\lambda_0)}\ {\ln^2\alpha\over \alpha}\ {\rm and}\  Y \gg {9\pi\over
16(1-\lambda_0)^3N_c\chi^{\prime\prime}(\lambda_0)}\ {\ln^2(\alpha Y)\over \alpha}. 
\end{displaymath}

\noindent The first condition is equivalent to $2\Delta(\alpha) \ll {\sqrt{t}},$ while the second is a consequence of replacing $\lambda_c$ by
$\lambda_0$ in (28).

Expression (47) exhibits a scaling behaviour, since the amplitude depends only on the ratio $Q^2/Q_c^2(Y)$; lines with constant $Q^2/Q_c^2(Y)$
will be lines of constant amplitude.  This in fact agrees with recent numerical solutions\cite{Mot} of the Balitsky-Kovchegov equation, where
it was found that the scaling behavior is extended in a domain outside the saturation region.

It is also interesting to notice that, even without the knowledge of the exact form of the non-linear effects, one is able to determine the
amplitude $T$ in (47), with no need to introduce any unknown parameters.

Finally, we come to the issue of the rapidity dependence of the saturation momentum.  In our approach, $\rho_s$ is defined in (43), and by using
$\rho_s=\ln(Q_s^2/\mu^2)$ we are lead to

\begin{equation}
Q_s^2(Y) = \mu^2\left[{\sqrt{\ln(1/\alpha)}} \alpha\right]^{2\over 1-\lambda_0} {\exp[{2\alpha N_c\over \pi}\ {\chi(\lambda_0)\over
1-\lambda_0}Y]\over (\alpha Y)^{3\over 2(1-\lambda_0)}},
\end{equation}

\noindent where of course an overall constant factor is free.

\section{The Running Coupling Case}

We now extend our main results, given in (47) and (48), to the case of BFKL evolution using a running coupling.  Surprisingly this turns out to
be not too difficult, although we shall not be able to get a result quite as complete as (47) in that the overall constant in our amplitude
will be undetermined even when $\ln (Q^2/Q_s^2) \gg 1.$

Eq.(1) is no longer a good starting point for running coupling BFKL evolution.  Rather, we write $T(Q,\mu,Y)$ in the general form

\begin{equation}
T(Q,\mu,Y) = \alpha(Q)\int {d\omega\over 2\pi i} \int {d\lambda\over 2\pi i} T_{\omega\lambda} \exp[\omega Y-(1-\lambda)(\rho-\rho_c) + \gamma\ 
\ln Y]
\end{equation}

\noindent where in this Section $\rho = \ln (Q^2/\Lambda^2),$ {{\rm not}\  $\ln (Q^2/\mu^2$)}, and where $\rho_c=\rho_c(Y)$ and
$\gamma$ will be specified in a moment.  $T_{\omega\lambda}$ has no $Q^2$ or $Y-$depdendence, but does contain the $\mu$-dependence of $T.\  
\Lambda$ is the usual QCD $\Lambda$-parameter and the $\omega$-integration in (49) goes parallel to the imaginary axis and to the right of any
singularities
$T_{\omega\lambda}$ may have in $\omega.$  We now view $T$ as a function of $\rho$ and $Y$ with the $\mu$-dependence suppressed.  Our
normalization is as in the fixed coupling case.  The BFKL equation is, schematically, 

\begin{equation}
{dT/\alpha(\rho)\over dY} = {\alpha(\rho)N_c\over \pi} K\ast(T/\alpha)
\end{equation}

\noindent where  $K$ is the usual BFKL kernel.  Eq.(50) is easily applied to (49) if one uses the fact that 

\begin{equation}
K\ast e^{-(1-\lambda)\rho} = 2\chi(\lambda) e^{-(1-\lambda)\rho}
\end{equation}

\noindent with

\begin{equation}
\chi(\lambda) = \psi(1) - {1\over 2} \psi(\lambda) - {1\over 2}\ \psi(1-\lambda)
\end{equation}

\noindent the usual BFKL eigenvalue function.  Using (50) and (51) on (49) gives

\begin{displaymath}
0=\int{d\omega d\lambda\over (2\pi i)^2} T_{\omega\lambda}\exp\left[\omega Y-(1-\lambda)(\rho-\rho_c)+\gamma\ln Y\right]
\end{displaymath}
\begin{equation}
\times \left\{\omega +
{d\rho_c\over dY}(1-\lambda)+{\gamma\over Y}-{2\alpha(\rho)N_c\over \pi} \chi(\lambda)\right\}.
\end{equation}

\noindent We now choose $\rho_c(Y)$ in such a way  that when $\rho=\rho_c(Y)$ the amplitude  $T$ becomes almost constant, thus following
closely our fixed coupling procedure of Sec.6.  This is done by choosing $\rho_c(Y)$ so that as much as possible of ${2\alpha N_c\over \pi}
\chi(\lambda)$ is cancelled by $(1-\lambda){d\rho_c\over dY}$ in the vicinity of the $\lambda-$values that dominate (49) and (53) when $\rho$
is near $\rho_c(Y).$  Suppose $\lambda-$values near $\lambda_c$ are dominant.  Then write

\begin{equation}
\chi(\lambda) \simeq \chi(\lambda_c) + (\lambda-\lambda_c)\chi^\prime(\lambda_c) + {1\over 2}
(\lambda-\lambda_c)^2\chi^{\prime\prime}(\lambda_c),
\end{equation}

\noindent in a ``diffusion approximation'' much like that introduced by Camici and Ciafaloni\cite{Cam}.  Furthermore, when $\rho$ is not too
far from $\rho_c$ write

\begin{equation}
\alpha(\rho) \simeq {1\over b\rho_c} (1-{\rho-\rho_c\over \rho_c}).
\end{equation}

\noindent By requiring

\begin{equation}
{d\rho_c\over dY} (1-\lambda_c) - {2N_c\over \pi b\rho_c} \chi(\lambda_c) + {\gamma\over Y} = 0
\end{equation}

\noindent and

\begin{equation}
{d\rho_c\over dY} + {2N_c\over \pi b\rho_c} \chi^\prime(\lambda_c) =0
\end{equation}

\noindent the leading parts of ${2\alpha N_c\over \pi} \chi(\lambda)$ are cancelled and the bracket in (53) becomes

\begin{align}
\{\ \}\simeq \bigg\{& \omega-{N_c\over \pi b\rho_c} \chi^{\prime\prime}(\lambda_c)(\lambda-\lambda_c)^2+{2N_c(\rho-\rho_c)\over \pi b\rho_c^2}\nonumber\\
& \times \left[\chi(\lambda_c)+ (\lambda-\lambda_c) \chi^\prime(\lambda_c) + {1\over 2 }
(\lambda-\lambda_c)^2\chi^{\prime\prime}(\lambda_c)\right]\bigg\}
\end{align}

\noindent for all terms of the type given in (54) and (55).  The final term in (58) is of the same form, in $(\lambda-\lambda_c)$ as the second
term, but smaller by a factor of ${\rho-\rho_c\over \rho_c},$ so we drop it in what follows.

If we write

\begin{equation}
T=\alpha(Q) \exp[-(1-\lambda_c)(\rho-\rho_c) + \gamma\ln Y]\psi(\rho,Y) T_0(\mu)
\end{equation}

\noindent then the factors of $(\lambda-\lambda_c)$ in (58) can be taken to be ${\partial\over \partial\rho}$ factors acting on $\psi.$ 
Eq.(53) then reads

\begin{equation}
\left\{{\partial\over \partial Y}-{N_c\chi^{\prime\prime}(\lambda_c)\over \pi b\rho_c}{\partial^2\over \partial\rho^2} + {2N_c(\rho-\rho_c)\over
\pi b\rho_c^2}[\chi(\lambda_c) + \chi^\prime(\lambda_c){\partial\over \partial\rho}]\right\}\psi=0
\end{equation}

\noindent which becomes our basic equation.  Of course for this whole procedure to work the integrals over $\lambda$ in (49) and (53) should be
dominated by values where $\lambda \simeq \lambda_c.$  We can check this at the end by verifying that ${\partial\over \partial\rho} \psi \lesssim
{1\over Y^{1/6}}\psi$ so that $\lambda-\lambda_c \simeq Y^{-1/6}$ in (49).  Before solving (60) we first turn to a determination of
$\rho_c(Y)$ from (56) and (57).

Comparing (56) and (57) to (25) and (26) one sees that (57) is a saddle-point condition while (56) is the condition that  $T$  change slowly,
depending on $\gamma,$ with $Y.$  Substituting (57) into (56) gives

\begin{equation}
\chi^\prime(\lambda_c)(1-\lambda_c) + \chi(\lambda_c) = {\pi b\rho_c\gamma\over 2N_cY}
\end{equation}

\noindent analogous to (27).  Anticipating that $\rho_c(Y)/Y \ll 1 $ we can expand about $\lambda_0,$ satisfying (6), to get

\begin{equation}
\lambda_c-\lambda_0 = {\gamma\pi b\rho_c\over 2 N_c\chi^{\prime\prime}(\lambda_0)(1-\lambda_0)Y}
\end{equation}

\noindent analogous to (29).

Now expand (57) about $\lambda_0$ keeping only terms up to first order in $\lambda_c-\lambda_0.$  One finds

\begin{equation}
{d\rho_c\over dY} + {2N_c\over \pi b\rho_c} \chi^\prime(\lambda_0) + {2N_c\over \pi b\rho_c} (\lambda_c-\lambda_0)
\chi^{\prime\prime}(\lambda_0) = 0
\end{equation}

\noindent Defining $\rho_0$ to be the solution to

\begin{equation}
{d\rho_0\over dY} + {2N_c\over \pi b\rho_0} \chi^\prime(\lambda_0) = 0,
\end{equation}

\noindent one gets

\begin{equation}
\rho_0(Y) = {\sqrt{{4N_c\over \pi b}{\chi(\lambda_0)\over 1-\lambda_0}(Y+Y_0)}}
\end{equation}

\noindent with $Y_0$ integration constant. Multiplying by $\rho_c,$ Eq.(63) can be written as

\begin{equation}
{d\over dY} \rho_c^2 = {4N_c\chi(\lambda_0)\over (1-\lambda_0)\pi b} - {2\gamma\rho_0\over (1-\lambda_0)Y}
\end{equation}

\noindent where we have replaced $\rho_c$ by $\rho_0$ in evaluating $\lambda_c-\lambda_0$ in (63).  Eq.(66) is easily integrated to give

\begin{equation}
\rho_c^2=\rho_0^2 - {4\gamma\rho_0\over 1-\lambda_0} + {\rm const}
\end{equation}

\noindent or

\begin{equation}
\rho_c(Y) = \rho_0(Y) - {2\gamma\over 1-\lambda_0}
\end{equation}

\noindent where we have dropped terms of size $1/{\sqrt{Y}}$ on the right-hand side of (68).  Eq.(68) is the generalization of (31) to the
running coupling situation.

Now we are in a position to solve (60), at least in an approximate way.  In the denominators in (60) we can replace $\rho_c$ by $\rho_0$ for
large $Y.$  Then dropping $Y_0$ in $\rho_0$ gives

\begin{equation}
[{\partial\over \partial Y} - {a\over {\sqrt{Y}}}{\partial^2\over \partial\rho^2} + {c(\rho-\rho_c)\over Y} - {\rho-\rho_c\over
2Y}{\partial\over \partial\rho}]\psi = 0,
\end{equation}

\noindent where

\begin{equation}
a = {\sqrt{{N_c(1-\lambda_0)(\chi^{\prime\prime}(\lambda_0))^2\over 4\pi b\chi(\lambda_0)}}}
\end{equation}

\noindent and

\begin{equation}
c = {1-\lambda_0\over 2}.
\end{equation}

\noindent Defining new variables

\begin{equation}
\xi = \left({c\over a}\right)^{1/3} {\rho-\rho_c\over Y^{1/6}}
\end{equation}

\noindent and

\begin{equation}
t = 6a^{1/3} c^{2/3} Y^{1/6},
\end{equation}

\noindent Eq.(69) becomes

\begin{equation}
[{\partial\over \partial t} - {\partial^2\over \partial\xi^2} + \xi - {4\over t} \xi {\partial\over \partial\xi}] \psi(\xi, t) = 0.
\end{equation}

We do not know how to solve (74) exactly.  However, we can find an approximate solution of the form

\begin{equation}
\psi = {1\over t^2}Ai(\xi-\lambda) \exp[-{\xi^2\over t} - \lambda t]
\end{equation}

\noindent where $\lambda$ is a constant and $Ai$ is the  Airy function.  One can easily check that (75) satisfies (74) up to terms of size
${\xi^2\over t^2}\psi.$  Since the Airy function decreases as $\exp[-{2\over 3}\xi^{3/2}]$ for large values of $\xi$ our ansatz satisfies (74)
in the region where $\psi$ is not exponentially small.  (Since we wish to use (75) over a wide range of values of   $t$  it is important that
(74) be satisfied including all terms of size $1/t, \xi/t$ and $\xi^2/t$ as such terms become important after integrating (73) over
$t.$  This is indeed the case and this requires the $\exp[-\xi^2/t]$ term in (75) whose actual {\underline {value}} is small in the region of
interest.)

Return to (59).  Using (75) and expressing all variables in terms of $\xi$ and $t$ one gets

\begin{equation}
T=\exp[-{\xi t\over 3} + (6\gamma-5) \ln t -\lambda t] Ai(\xi-\lambda) \tilde{T}_0
\end{equation}

\noindent where we have replaced ${1-\lambda_c\over 1-\lambda_0}$ by $1$ in the exponent, and where some constants have been included with $T_0$
to give $\tilde{T}_0.$  Following our procedure in the fixed coupling case we must impose the condition that $T$ not become large in the
saturation region.  Since the first term in (76) grows strongly for negative value of $\xi$ we should choose $\lambda$ so that the maximum of
$T$ has a value on the order of $1/\mu^2$ (see 44) at $\xi = \xi_s$ and then goes to zero at $\xi = \xi_1+\lambda,$\  a value slightly less than
$\xi_s,$ where $\xi_1$ is the first zero of $Ai(\xi).$  To that end write

\begin{equation}
\xi = \lambda + \xi_1 + \delta \xi.
\end{equation}

\noindent Then

\begin{equation}
Ai(\xi -  \lambda) \simeq Ai^\prime(\xi_1) \delta\xi
\end{equation}

\noindent leading to

\begin{equation}
T=\exp[-{1\over 3}(\xi_1+4\lambda) t-{1\over 3}t\delta\xi + \ln \delta\xi + (6 \gamma -5) \ln t]Ai^\prime(\xi_1)\tilde{T}_0. 
\end{equation}

\noindent $T$ has a maximum at

\begin{equation}
\delta\xi = 3/t
\end{equation}

\noindent at which the value of $T$ is

\begin{equation}
T_{max}=\exp[-{1\over 3}(\xi_1+4\lambda) t+\ln (3/e) + 6(\gamma -1) \ln t]Ai^\prime(\xi_1) \tilde{T}_0.
\end{equation}

\noindent Clearly we must take $\gamma = 1$ and $\lambda = - \xi_1/4$ in order that $T_{max}$ be independent of $Y.$

Defining
\begin{equation}
\rho_s=\left({a\over c}\right)^{1/3} Y^{1/6}[\xi_1+ \lambda + 3/t] + \rho_c
\end{equation}

\noindent one finds

\begin{equation}
\rho_s={\sqrt{{4N_c\over \pi b}{\chi(\lambda_0)\over 1-\lambda_0} Y}} + {3\over 4}\left({a\over c}\right)^{1/3} \xi_1Y^{1/6} - {1\over
1-\lambda_0}.
\end{equation}

\noindent Rewriting (79) in terms of the more physical variables $\rho$ and $Y$ gives

\begin{equation}
T=e^{-(1-\lambda_0)(\rho-\rho_s)}Y^{1/6}Ai\left(\xi_1+\left({c\over a}\right)^{1/3}{1\over Y^{1/6}}[\rho-\rho_s+{1\over
1-\lambda_0}]\right)\tilde{T}_0^\prime
\end{equation}

\noindent or

\begin{equation}
T=\left({Q^2\over Q_s^2}\right)^{-(1-\lambda_0)}\left[\ln {Q^2\over Q_s^2} + {1\over 1-\lambda_0}\right]T_0^\prime
\end{equation}

\noindent when $\ln (Q^2/Q_s^2) + {1\over 1-\lambda_0}\ll Y^{1/6}.\tilde{T}_0, T_0^\prime$ and $\tilde{T}_0^\prime$ are all related by rather
trivially calculated constant factors.  Eq.(85) is very close to (47), however in the present situation we have no control over the value of
$T_0^\prime$ in contrast to the fixed coupling case.

It was the choice of $\gamma = 1$ in (76) which allowed us to cancel the factor of $\alpha(Q) \sim 1/{\sqrt{Y}}$ present as a prefactor in the
right-hand side of (49).  That is, we have complete control over the $Y$ and $Q^2$-dependence of $T$ near saturation, but we have lost our
ability to deal with the $\mu-$dependence.  Referring to (83) it would appear that there are no free factors left in determining $Q_s,$ apart
from an uninteresting possibility of a not large additive constant associated with our choice of defining $\rho_s$ at exactly the maximum of
$T.$  In particular it seems that $\rho_s,$ or $Q_s,$ is completely independent of $\mu$ in contrast to the fixed coupling case where $\mu$
set the scale $Q_s.$  Indeed, we believe  this to be the case, and this makes the present discussion valid also for the scattering of a dipole
of size $1/
Q$ on a proton, however, one must be careful as to the range of $Y$ in which (83)-(85) can be applied.  To see the limits on $Y$ we suppose that
$\mu/\Lambda \gg 1,$ then for $Y$ not too large there should be a range of $Y$ for which the fixed coupling description is applicable even when
running coupling effects are allowed\cite{Cia}.  Referring to (48) one easily sees that so long as 

\begin{equation}
\ln (\mu^2/\Lambda^2) \gg {2\alpha(\mu)N_c\over \pi} {\chi(\lambda_0)\over 1-\lambda_0}Y
\end{equation}

\noindent $\alpha(\mu) - \alpha(Q_s) \ll \alpha(\mu),$ and the fixed coupling approach should be valid.  This suggests that the boundary where
one must change from a fixed coupling to a running coupling description occurs at a transition value of $Y$

\begin{equation}
Y_{trans}\simeq {\pi(1-\lambda_0)\over 2bN_c\chi(\lambda_0)}{1\over \alpha^2(\mu)}.
\end{equation}

\noindent  If $Y \ll Y_{trans}$ the fixed coupling decription is applicable while the running coupling description of this section is
applicable when $Y\gg Y_{trans}.$  In order to determine $\tilde{T}_0^\prime$ in (84), or $T_0^\prime$ in (85) we would have to follow the
$Y$-evolution of the system through the transition rapidity, a task which we have not attempted.

\vskip 10pt

\appendix 
\noindent{\bf Appendix A}
\vskip 10pt

Here we calculate the amplitude in the non-forward case following\cite{Nav,haw}.  With the definitions and notation used in Section
4, we start by exchanging the denominators of $E_q^{0\nu}(x)$ as given in (12), by introducing a Feynman parameter $\beta.$  After integrating
over $\vec{R}$ we find

$$E_q^{0\nu}(x) =  {4i\nu\over \pi} (2q)^{2i\nu}{\Gamma^2(1+i\nu)\over \Gamma^2(1+2i\nu)} e^{i\vec{q}\cdot \vec{x}/2}$$

$$ \times  \int_0^1{d\beta\over{\sqrt{\beta(1-\beta)}}}e^{-i\beta\vec{q}\cdot \vec{x}} K_{2i\nu}({\sqrt{\beta(1-\beta)}}\ qx)\eqno{(A.1)}$$

\noindent For small $xq$ the exponentials can be approximated by 1 and expanding the Bessel function for small argument as

$$K_{2i\nu}\left({\sqrt{\beta(1-\beta)}}\ qx\right)= {1\over 4i\nu}\left\{\Gamma(1+2i\nu)\left[{2\over
qx{\sqrt{\beta(1-\beta)}}}\right]^{2i\nu}-c.c.\right\}\eqno{(A.2)}$$

\noindent where $c.c.$ stands for complex conjugate, one finds

$$E_q^{0\nu}(x) = {1\over \pi}(2q)^{2i\nu}{\Gamma^2(1+i\nu)\over \Gamma^2(1+2i\nu)}$$

$$\times [\Gamma(1+2i\nu)({2\over qx})^{2i\nu}\int_0^1d\beta\left({\sqrt{\beta(1-\beta)}}\ \right)^{-1-2i\nu}-c.c.].\eqno{(A.3)}$$

\noindent The integral over the Feynman parameter $\beta$ is straightforward and we are lead to

$$E_q^{0\nu}(x)=(2q)^{2i\nu}{\Gamma(1-2i\nu)\over \Gamma(1+2i\nu)}\left[{\Gamma^2(1+i\nu)\over \Gamma^2(1-i\nu)}\left({8\over
qx}\right)^{2i\nu}-\left({qx\over 8}\right)^{2i\nu}\right].\eqno{(A.4)}$$

\noindent The factor outside the bracket in the last expression is just a phase, when $\nu$ is real, and therefore cancels when we consider the
product $E_q^{0\nu\ast}(x_0)E_q^{0\nu}(x),$ for which we have

$$E_q^{0\nu\ast}(x_0)E_q^{0\nu}(x)=2\left({x_0\over x}\right)^{2i\nu}-2\ {\Gamma^2(1-i\nu)\over \Gamma^2(1+i\nu)}\left({q^2x x_0\over
64}\right)^{2i\nu},\eqno{(A.5)}$$

\noindent plus terms odd in $\nu$ which will vanish when the $\nu$ integration is done.  The second term may be written as

$$
-2\left({\Gamma(1-i\nu)\over \Gamma(1+i\nu)}e^{-2\gamma i\nu}\right)^2\left({e^{2\gamma}q^2x x_0\over 64}\right)^{2i\nu} = $$
$$-2\left(1-{4i\zeta(3)\over 3}\nu^3+\cdot\cdot\cdot\right)\left({e^{2\gamma}q^2x x_0\over 64}\right)^{2i\nu},\eqno{(A.6)}$$

\noindent so that, neglecting terms of order $\nu^3,$ we finally have

$$E_q^{0\nu\ast}(x_0)E_q^{0\nu}(x) = 2\left({x_0\over x}\right)^{2i\nu}-2\left(cq^2 x x_0\right)^{2i\nu},\eqno{(A.7)}$$

\noindent with $c=e^{2\gamma}/64.$  Now the $\nu$ integation in  (10) can be done in the standard way; by expanding $\chi(\nu)$ around
$\nu=0$, keeping only up to quadratic terms and performing the Gaussian integration.  This leads to Eq.(13).  The second term in (A.7) gives
rise to the second term in (13).

\end{document}